\documentclass[runningheads]{llncs}
\usepackage{amsmath,amssymb,amsfonts}
\usepackage{textcomp}
\usepackage[pdftex]{graphicx}
\usepackage[misc]{ifsym}
\usepackage[linesnumbered,ruled]{algorithm2e}
\usepackage{color, colortbl}
\usepackage{spverbatim}
\usepackage{etoolbox}
\usepackage{multirow}
\usepackage{booktabs}
\usepackage{bm}
\usepackage{bbm}
\usepackage{enumitem}
\usepackage{bbding}
\usepackage{xcolor}
\usepackage{soul}
\usepackage{bbding}
\usepackage{hyperref}
\usepackage{url}
\hypersetup{
    colorlinks=true,
    linkcolor=blue,
    filecolor=magenta,      
    urlcolor=red,
}

\usepackage[utf8]{inputenc}

\begin{document}
\title{SpecTr: Spectral Transformer for Hyperspectral Pathology Image Segmentation}
%
\titlerunning{SpecTr: Spectral Transformer}
%
%
%
\author{Boxiang Yun\inst{1} \and
Yan Wang\inst{1}\Envelope \and
Jieneng Chen\inst{2} \and
Huiyu Wang\inst{2} \and\\
Wei Shen\inst{3} \and 
Qingli Li\inst{1}}
\authorrunning{B. Yun \emph{et al.}}
\institute{$^{1}$East China Normal University $^{2}$Johns Hopkins University \\ $^{3}$Shanghai Jiaotong University \\
{\tt\small wyanny.9@gmail.com}\\
}
\maketitle              
\begin{abstract}
Hyperspectral imaging (HSI) unlocks the huge potential to a wide variety of applications relied on high-precision pathology image segmentation, such as computational pathology and precision medicine. Since hyperspectral pathology images benefit from the rich and detailed spectral information even beyond the visible spectrum, the key to achieve high-precision hyperspectral pathology image segmentation is to felicitously model the context along high-dimensional spectral bands. Inspired by the strong context modeling ability of transformers, we hereby, for the first time, formulate the contextual feature learning across spectral bands for hyperspectral pathology image segmentation as a sequence-to-sequence prediction procedure by transformers. To assist spectral context learning procedure, we introduce two important strategies: (1) a sparsity scheme enforces the learned contextual relationship to be sparse, so as to eliminates the distraction from the redundant bands;
(2) a spectral normalization, a separate group normalization for each spectral band, mitigates the nuisance caused by heterogeneous underlying distributions of bands. 
We name our method Spectral Transformer (SpecTr), which enjoys two benefits: (1) it has a strong ability to model long-range dependency among spectral bands, and (2) it jointly explores the spatial-spectral features of HSI. Experiments show that SpecTr outperforms other competing methods in a hyperspectral pathology image segmentation benchmark without the need of pre-training. Code is available at~\url{https://github.com/hfut-xc-yun/SpecTr}.

\end{abstract}
\section{Introduction}

Hyperspectral imaging (HSI) is a technique that analyzes how a wide spectrum of light interacts with observed materials, measuring the amount of light that is emitted, reflected or transmitted from a target \cite{Paoletti2019deep}, instead of assigning primary colors (red, green, blue) to each pixel. Detailed spectral information is presented in hundreds of narrow and contiguous spectral bands. Since different molecules' responses to light are different, utilizing HSI, spectral information across bands acquires biochemical properties invisible to the naked eye from both stained and unstained histological specimens, which brings opportunities for digital and computational pathology and precision medicine.

With the surge of deep neural networks, hyperspectral pathology image segmentation is undergoing a transition from traditional machine learning methods \cite{Lu2014toward} to deep learning based methods \cite{Ortega2020hyperspectral}. Traditional methods designed spatial feature extraction methods such as watershed, wavelet, local binary pattern, and spectral feature mapping methods such as spectral unmixing, band selection. But these methods suffer from the low efficiency of feature fusion, limited representation learning ability. High-dimensional spectral bands in HSIs are highly correlated \cite{Chang1999a}, exploring the wide range of band relationships is critical to achieve high-precision hyperspectral pathology image segmentation results. Deep learning, especially the u-shape architecture, has become the de-facto standard and achieved tremendous success for various medical image segmentation tasks. However, u-shape networks, no matter 2D or 3D, have limited ability to model long-range context along the rich spectral dimension, which restricts these methods from achieving good performance on hyperspectral pathology image segmentation \cite{Wang2021identification,Trajanovski2020tongue}. 

Inspired from transformers' strong ability in modeling long-range context \cite{Dosovitskiy2021an,Zheng2020rethinking,Chen2021TransUNet,Maria2021medical}, we hereby, for the first time, formulate context-aware spectral band representation learning as a sequence-to-sequence prediction problem, and instantiate it by transformers. Moreover, to assist the spectrum context modeling process, two important schemes are introduced. {First, since standard transformers learn dense correlations among all bands, the learned contextual representing may be polluted by irrelevant bands. We thereby introduce an important sparsity scheme \cite{Correia2019adaptively} into transformers to ensure the sparsity of the learned correlations.} Second, a spectral normalization, a separate group normalization for each band, is proposed to ease the nuisance caused by heterogeneous underlying distributions of bands. Our transformers are embedded into the encoder part of a u-shape segmentation network, with the input of convolutional features at each spatial location. 
{The convolutional features are equipped with spatial information, which are then entangled with spectral information encoded by transformers, forming a \emph{joint spatial-spectral} representation learning framework.} 
{We term the proposed method as Spectral Transformer (SpecTr). Experimental results show that the two introduced schemes are vital for spectral contextual representation learning. Our SpecTr achieves significant better hyperspectral pathology segmentation results than other competitors, including CNN-based self-attention segmentation methods.}

\section{Related Work}

\textbf{Hyperspectral medical image segmentation}
Traditional hyperspectral medical image segmentation methods are mainly dedicated to extract hand-designed spectral and spatial features from hyperspectral data, and fuse features from the two domains \cite{Lu2014toward}. Deep learning methods become ubiquitous in the field of medical image segmentation, which also benefit hyperspectral medical image segmentation. 2D CNN is explored to segment tongue tumor from HSIs, with an additional layer to select important bands from HSIs \cite{Trajanovski2020tongue}. Since HSIs are 3D volumes, it is intuitive to use 3D CNNs to simultaneously learn spatial-spectral information. Hyper-Net \cite{Wang2021identification} first extracted 16 informative bands from an image, then fed the image into a dual-path module 3D CNN, which achieved promising results for hyperspectral pathology image segmentation.

\noindent\textbf{Transformer}
Transformer \cite{Vaswani2017attention} was first proposed for NLP tasks, it becomes state-of-the-art methods in some computer vision tasks \cite{Dosovitskiy2021an,Zheng2020rethinking,Parmar2018image}. Transformer, based solely on attention mechanisms, is famous for its sequence modeling ability. Parmar \emph{et al.} \cite{Parmar2018image} performed conditional image generation with the Image Transformer, and the generated images look more natural than other methods. ViT \cite{Dosovitskiy2021an} showed that applying a pure transformer to sequences of image patches can achieve favorable performance in image classification tasks. SETR \cite{Zheng2020rethinking} leverages merits from both transformer and CNN, where transformer is served as the encoder while CNN is the decoder. Very recently, TransUNet \cite{Chen2021TransUNet}, which also combines transformer and CNN, obtaining promising results for some medical segmentation tasks. Our SpecTr is the first model to learn the sequence representation of spectra for hyperspectral pathology image segmentation, and explores the context along spectral dimension which is preferable to segmentation.

\begin{figure}[t]
\begin{center}
    \includegraphics[width=1\linewidth]{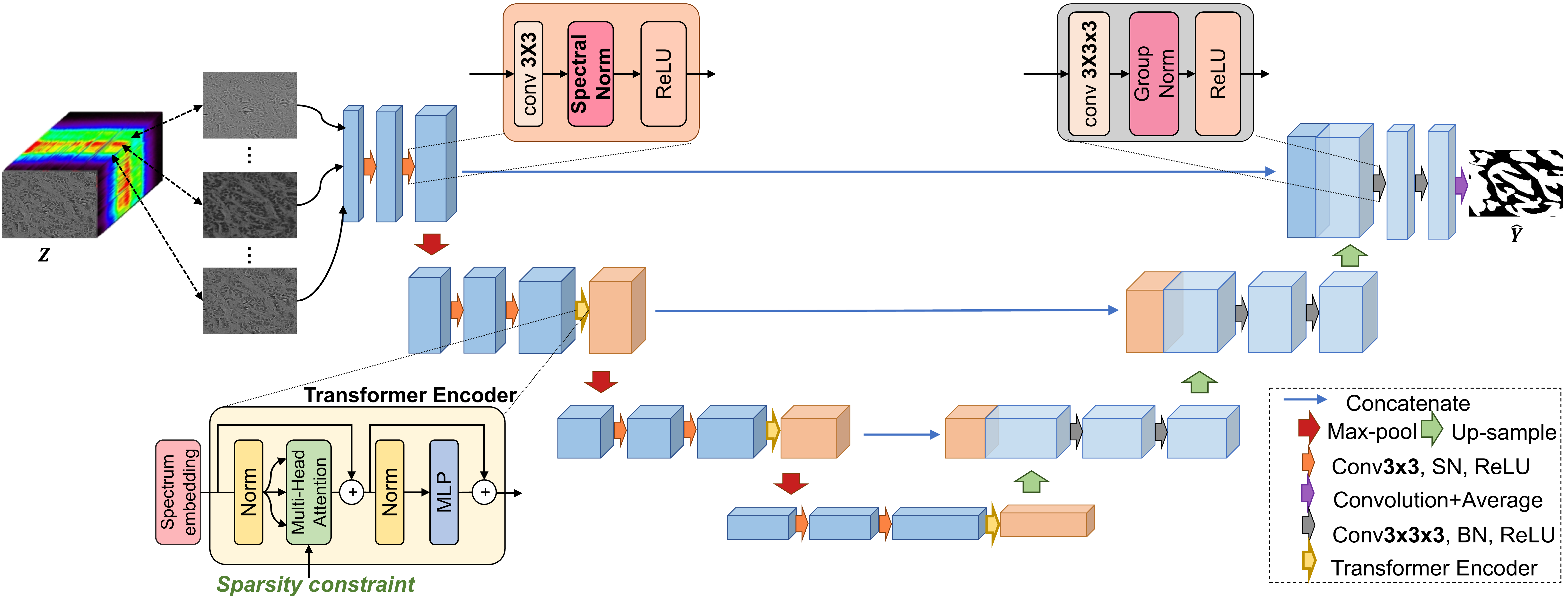}
\end{center}
\caption{{The architecture of SpecTr.} Follow the u-shape architecture, our SpecTr employs 2D convolutional kernel, spectral norm, and transformers with the sparsity constraint in the encoder. The decoder and skip connections are the same as 3D U-Net.
} 
\label{Fig:framework}
\end{figure}

\section{Methodology}
Mathematically, let $\mathbf{Z}\in\mathbb{R}^{W\times H\times L}$ denote the 3D volume of a hyperspectral pathology image,
where $W\times H$ is the spatial resolution, and $L$ is the number of spectral bands. The goal of image segmentation is to predict the per-pixel label map $\hat{\mathbf{Y}}\in\{0,1\}^{W\times H}$, indicating where the target is in $\mathbf{Z}$. Let $\mathcal{D}=\{\mathbf{Z}_i,\mathbf{Y}_i\}_{i=1}^N$ be our training set, where $\mathbf{Y}_i$ denotes the per-pixel annotation for image $\mathbf{Z}_i$. An overview of the model, which follows the u-shape architecture, is depicted in Fig.~\ref{Fig:framework}. Image $\mathbf{Z}$ is first decomposed into a sequence of spectral images in order of wavelength. Then the sequence of spectral images is fed into an alternating process of depth-wise convolution, spectral normalization and transformers with sparsity constraint to {produce a sequence of spatial-spectral contextual feature maps with the same sequence length}. Since the first resolution layer lacks semantic feature for each spectral location, transformers are added after the second, third and fourth resolution layers. Then the {sequence of feature maps are concatenated and then gradually} decoded {to generate the segmentation map}. Skip connections are employed, which from layers of equal resolution in the encoder retain essential high-resolution features to the decoder. 

\subsection{Spectral Normalization}
\begin{figure}[t]
\begin{center}
    \includegraphics[width=0.7\linewidth]{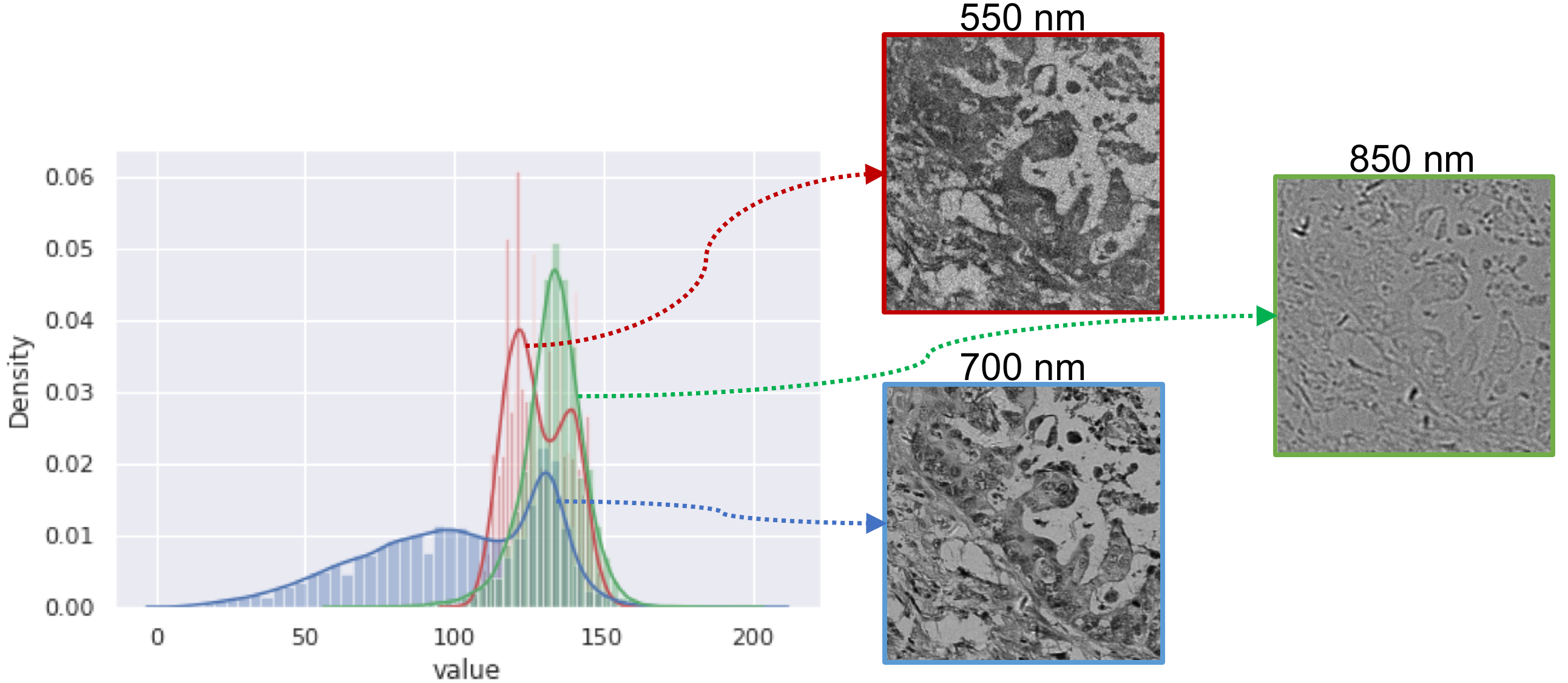}
\end{center}
\caption{Probability density functions (PDF) of different spectral images. Each PDF on the left is plotted by average five hyperspectral pathology images. Example spectral images of different wavelengths from one HSI are shown on the right.
} 
\label{Fig:SN}
\end{figure}
In the encoder, the decomposed spectral images are fed into depth-wise convolution \cite{Sandler2018mobilenet}, which applies a single filter to each input spectral image. Depth-wise convolution can learn spectrum embedding while keeping spectrum independent. A normalization processing is usually applied after convolution for reducing internal covariate shift, and therefore helping improve feature discrimination capability. Different spectral images have heterogeneous distributions, as shown in Fig.~\ref{Fig:SN}. When training deep networks that directly integrate all heterogeneous spectral images, we hypothesize this distribution mismatch among different spectral images adversely affects context learning among bands, and causes performance degradation. 

To this end, we propose Spectral Normalization (SN) to address this problem.
More specifically, let $\mathbf{F}(\mathbf{Z};\mathbf{\Theta})\in\mathbb{R}^{W'\times H'\times L'\times C}$ denote the feature map produced after a certain depth-wise convolutional layer, where $C$ is the number of feature channels, and $\mathbf{\Theta}$ is the parameter of all previous layers. We assign an individual group normalization layer for $\mathbf{f}_{s}(\mathbf{Z};\mathbf{\Theta})\in\mathbb{R}^{W'\times H' \times C}$, which is the feature map at spectral location ${s}$, to effectively tackle the inter-spectrum discrepancy.

\subsection{Transformer with Sparsity Constraint}
\textbf{Spatial-spectral Embedding} The feature vector $\mathbf{h}_{\mathbf{x},{s}}(\mathbf{Z};\mathbf{\Theta})\in\mathbb{R}^{C}$ at each spatial location $\mathbf{x}$ and spectral location ${s}$ on the feature map $\mathbf{F}(\mathbf{Z};\mathbf{\Theta})$ represents a spatial-spectral feature element. The feature element is learned by 2D spatial convolution, and will be fed into a transformer to learn spectral context. Followed by \cite{Dosovitskiy2021an,Zheng2020rethinking}, a linear projection matrix $\mathbf{E}\in\mathbb{R}^{C\times D}$ is learned to map the feature element $\mathbf{h}_{\mathbf{x},{s}}$ into a latent $D$-dimensional embedding space. We omit parameters for notational simplicity. To encode the spectral location information, we learn a specific embedding $p_{s}$ for every location $s$, which is added to the spatial-spectral embedding to form the sequential input: $\mathbf{z}_0=[\mathbf{h}_{\mathbf{x},1}\mathbf{E}+p_1; \mathbf{h}_{\mathbf{x},2}\mathbf{E}+p_2; ...; \mathbf{h}_{\mathbf{x},L'}\mathbf{E}+p_{L'}]$

\noindent{\textbf{Transformer}} The transformer consists of multiheaded self-attention (MSA), MLP blocks, layer normalization (LN) and residual connections. The output of the transformer can be computed by:
\begin{align}
    \mathbf{z}'_1 &= \text{MSA}(\text{LN}(\mathbf{z}_0))+\mathbf{z}_0,\\
    \mathbf{z}_1 &= \text{MLP}(\text{LN}(\mathbf{z}'_1))+\mathbf{z}'_1.
\end{align}

\noindent{\textbf{Transformer with Sparsity}} The attention distribution of each head in MSA is predicted typically using the softmax normalization function, which results in non-zero weights for all context bands. Our purpose is to learn a meaningful context among bands for segmentation tasks, \emph{i.e.}, to learn useful bands and get rid of noisy or redundant bands. Instead of using softmax, we employ $\alpha$ entmax as proposed in \cite{Correia2019adaptively}:
\begin{align}
    \text{Att}(\mathbf{Q}&,\mathbf{K},\mathbf{V}) = \boldsymbol{\pi}\left(\frac{\mathbf{Q}\mathbf{K}^{\top}}{\sqrt{d}}\right)\mathbf{V}, \\
    \boldsymbol{\pi}(\mathbf{X}&)_{ij} = \alpha\text{ -entmax}(x_i)_j.
\end{align}
If $\alpha=1$, it is exactly softmax mapping. If $\alpha>1$, it moves away from softmax towards sparse mappings. If $\alpha=2$, a complete sparse mappings are obtained. We set randomly-initialized $\alpha$ values for all heads and the values oscillate between $1$ and $2$ by jointly optimized with other parameters of the network.

\section{Experimental Results}
\subsection{Experimental Setup}
\textbf{Dataset}: We use multi-dimensional choledoch dataset for cholangiocarcinoma diagnosis \cite{Zhang2019a}, where $514$ scenes with high quality labels are selected for training and testing. When using microscopy hyperspectral imaging system to capture images, the light transmitted from the choledoch tissue slice was collected by the microscope with the objective lens of $20\times$. The wavelength is from $550$ nm to $1000$ nm, which ends up with $60$ spectral bands for each scene. The image size of a single band image of the microscopy hyperspectral date cube is $1280\times 1024$. We randomly split all scenes into $401$ for training and $113$ for testing.

\noindent\textbf{Implementation Details and Evaluation Metric}: Similar as \cite{Wang2021identification}, since the entire hyperspectral pathology image is too large to be fed into the model, each is downsampled four times spatially, divided into $196\times 196\times 60$ image cubes. 
For data augmentation, we adopt online rotation whose degree is less than $90^\circ$ and vertical or horizontal flipping with probability of $0.2$ during training for all the methods. We set the learning rate of each parameter group using a cosine annealing schedule, with the initial learning rate to be $0.0003$, and weight decay is $5\times 10^{-4}$. We use Adam optimizer. The batch size is $1$, and the maximum number of training epoch is $75$ unless otherwise specified. Experiments are conducted using a single Nvidia Tesla V100-PCIE. \textbf{All components in our SpecTr do not need pre-training.} Multiple evaluation metrics, including Dice-S{\o}rensen similarity coefficient (DSC), Intersection of Union (IoU), Hausdorff Distance (HD) are computed for evaluation purpose.

\subsection{Comparison between SpecTr and Other Methods}

\begin{table}[t]
\setlength{\tabcolsep}{7pt}
\footnotesize
\centering
\caption{Performance comparison in DSC (\%), IoU (\%), Hausdorff Distance (HD) on cholangiocarcinoma segmentation. \textbf{Bold} denotes the best results for each metric.}
\label{Tab:comparison}
\begin{tabular}{lccc}
\toprule[0.15em]
Method & Mean DSC (Median, Max, Min) $\uparrow$ & IoU $\uparrow$ & HD $\downarrow$ \\
\midrule
HSI UNet \cite{Trajanovski2020tongue} & 62.92 (66.71, 99.93, 12.59) & 49.48 & 41.31\\
HSI Hyper-net \cite{Wang2021identification} & 69.74 (75.05, 99.95, 4.185) & 56.91 & 36.51\\ 
3D UNet \cite{Cicek20163d} & 72.19 (74.98, 99.95, 7.548) & 59.37 & 35.37\\ 
2D UNet \cite{Ronneberger2015unet} & 65.80 (66.52, 93.87, 17.97) & 50.91 & 36.46\\
Attn UNet \cite{Oktay2018attention} & 71.05 (73.38, 98.43, 19.01) & 57.46 & 34.32\\
UNet++ \cite{Zhou2018unet++} & 68.63 (70.70, 96.26, 17.09) & 54.45 & 35.95 \\
SpecTr (Ours) & \textbf{75.21} (77.92, 99.48, 16.44) & \textbf{62.44} &\textbf{31.60}\\
\bottomrule[0.15em]
\end{tabular}
\end{table}
We conduct comparisons between SpecTr and six competitors: 1) HSI UNet \cite{Trajanovski2020tongue}, 2) HSI Hyper-net \cite{Wang2021identification}, 3) 3D UNet \cite{Cicek20163d}, 4) 2D UNet \cite{Ronneberger2015unet}, 5) Attention UNet \cite{Oktay2018attention}, and 6) UNet++ \cite{Zhou2018unet++}. The first two competitors are state-of-the-arts for medical HSI segmentation. HSI UNet incorporated a spectral selection layer in the network to select a subset of spectra that leaded to optimal performance. HSI Hyper-net first adopted a spectrum selection module to select the most important bands out of all bands, and then trained a 3D Hyper-net to conduct segmentation. We follow the same settings as illustrated in the paper \cite{Trajanovski2020tongue,Wang2021identification}. The other four methods are de-facto standards for various medical image segmentation tasks. For 2D networks, we simply feed all spectral bands as the input channels to train the networks.

\begin{figure}[t]
\begin{center}
    \includegraphics[width=0.72\linewidth]{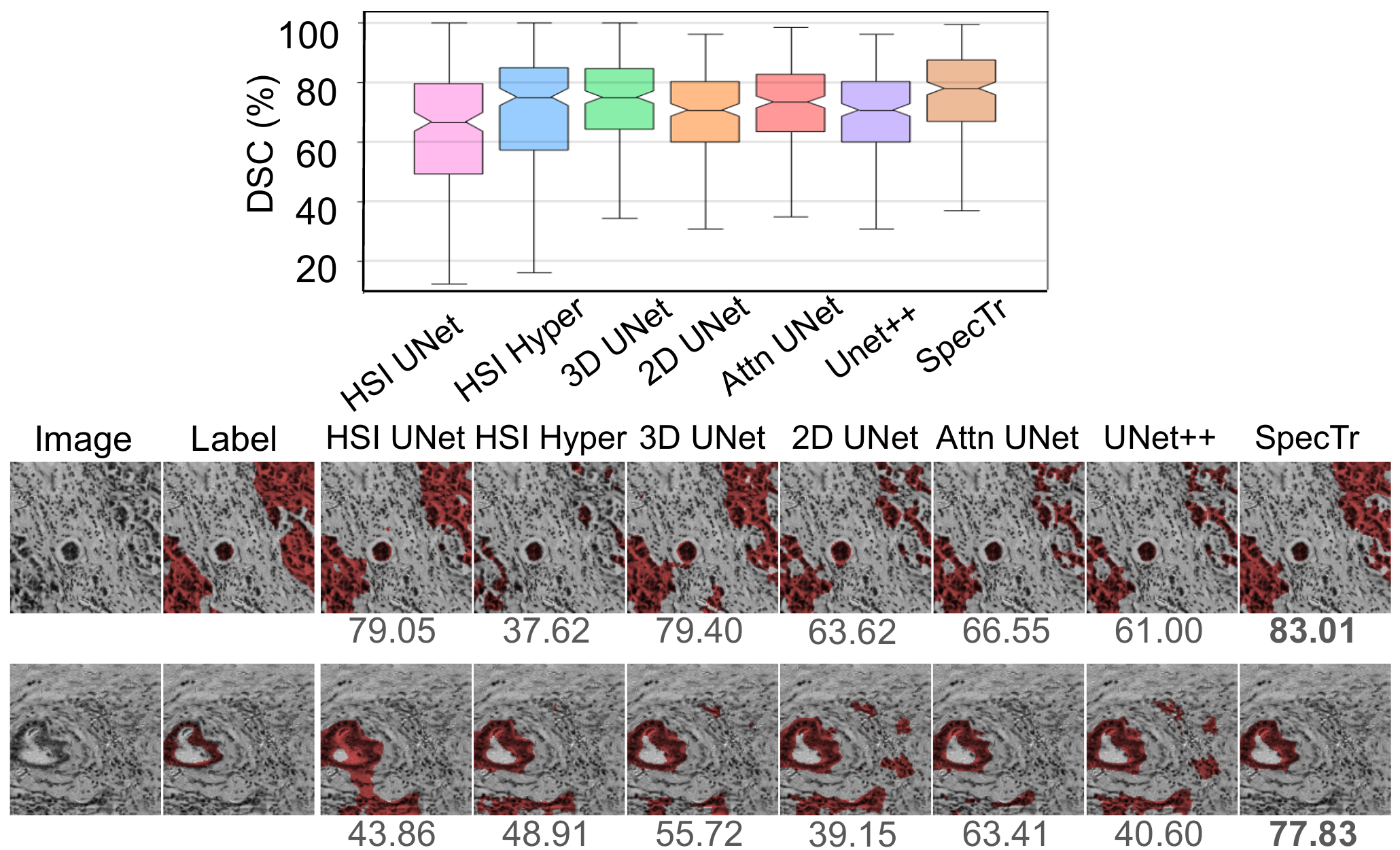}
\end{center}
\vspace{-0.5em}
\caption{Top: box plots on DSCs of SpecTr and all competing methods. Bottom: qualitative comparisons of cholangiocarcinoma segmentation. The spectrum of the image is 625 mn. Numbers on the bottom are segmentation DSCs ($\%$).
} 
\label{Fig:vis}
\vspace{-1em}
\end{figure}

Results are summarized in Table~\ref{Tab:comparison}. Our SpecTr performs much better than other competitors w.r.t. all evaluation metrics. In particular, SpecTr outperforms the two state-of-the-art hyperspectral medical image segmentation networks, \emph{i.e.}, HSI UNet and HSI Hyper-net by $12.29\%$ and $5.47\%$ in DSC. To make a fair comparison with the popular medical image segmentation networks, we change their batch normalization into group normalization. The 3D UNet with batch normalization obtains only $47.35\%$ in DSC. We also change 2D convolution in Attention UNet to 3D convolution, and obtains $68.28\%$ in DSC. The top figure in Fig.~\ref{Fig:vis} shows comparison results by box plots. We also illustrate segmentation results in the bottom figure of Fig.~\ref{Fig:vis} for qualitative comparison. We can see that compared with other methods, SpecTr can output more accurate segmentation results, which are more robust to the complicated background.

\subsection{Discussion on Spectral Band Context}
\begin{figure}[t]
\begin{center}
    \includegraphics[width=0.9\linewidth]{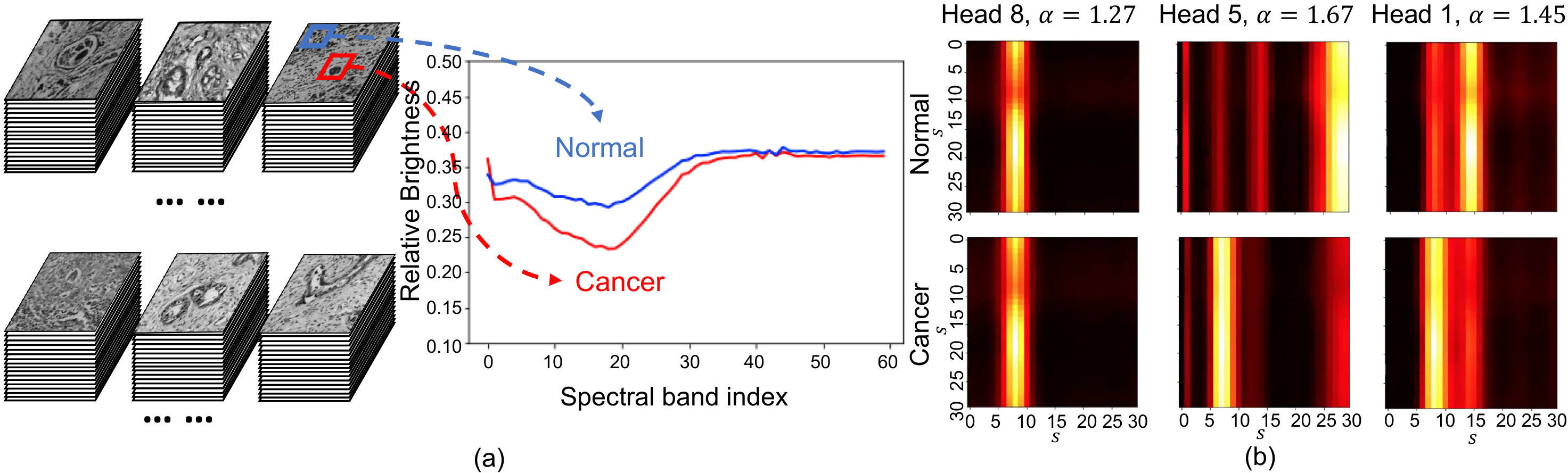}
\end{center}
\vspace{-0.5em}
\caption{(a) Plots of the intensity values versus spectral band index for ALL cancer and normal regions in the testing set. (b) Visualization of learned attention heads.
} 
\label{Fig:attention}
\end{figure}

Fig.~\ref{Fig:attention} (a) plots the average intensity values versus spectral band index for all cancer and normal regions in the testing set. The average intensity shows a discriminative region around the $20$th spectral band. In Fig.~\ref{Fig:attention} (b), we visualize a subset of the averaged attention heads which are sparse in the first transformer (after down-sampling the spectrum by a factor of $2$). In particular, head $5$ in the encoder self-attention layer becomes one of the sparsest head. Transformer successfully learns the attention spectra, \emph{i.e.}, the $5$-$10$ spectral locations in normal-head8, cancer-head8 and cancer-head5 (corresponding to $10$-$20$ spectral bands); $13$-$17$ spectral locations in normal-head1 ($26$-$34$ spectral bands). These two strongly correlated clusters are discriminative for segmentation. Interestingly, in normal-head5, the $25$-$30$ spectral locations ($50$-$60$ spectral bands) are also important, which suggests that though the intensity values within this range is similar, they may still contain discriminative information. The final segmentation masks are obtained via averaging probability maps from all spectral images (see the purple arrow in Fig.~\ref{Fig:framework}), which enforces the above-mentioned attention heads in transformers to project all queries into the discriminative keys. Thus, it leads to the vertical patterns in attention heads. 

\subsection{Ablation Study}
\begin{table}[t]
\renewcommand\arraystretch{0.3}
\setlength{\tabcolsep}{10pt}
\footnotesize
\centering
\caption{{{Ablation on the design of SpecTr.}}}
\label{Tab:ablation}
\begin{tabular}{cccccc|c}
\toprule[0.15em]
\multirow{2}{*}{Depthwise Conv} &\multicolumn{3}{c}{Transformer}&\multirow{2}{*}{Sparsity} &\multirow{2}{*}{SN} & \multirow{2}{*}{DSC ($\%$) }\\
\cmidrule(lr){2-4}
 & E2 & E3 & E4 &  & \\
\midrule
\checkmark & \checkmark & \checkmark & \checkmark & \checkmark & \checkmark & \textbf{75.21} \\ \rowcolor{gray!40}
  & \checkmark & \checkmark & \checkmark  & \checkmark & \checkmark & 72.79 \\
\checkmark &  &  & \checkmark & \checkmark & \checkmark & 72.17 \\ \rowcolor{gray!40}
\checkmark &  & \checkmark &  & \checkmark & \checkmark & 72.39 \\
\checkmark & \checkmark &  &  & \checkmark & \checkmark & 73.21 \\ \rowcolor{gray!40}
\checkmark & \checkmark & \checkmark & \checkmark & \checkmark &  & 70.66 \\ 
\checkmark & \checkmark & \checkmark & \checkmark &  & \checkmark & 73.28 \\\rowcolor{gray!40}
\checkmark  &  &  &  &  &  & 70.40 \\
\bottomrule[0.15em]
\end{tabular}
\vspace{-0.5em}
\end{table}
We conduct ablation experiments to analyze the influence of different designs and components for SpecTr. As shown in Table~\ref{Tab:ablation}, w/o depth-wise convolution means we replace depth-wise convolution with 3D convolution in SpecTr. Transformer (E2-E4) indicates whether the transformer is added after the second, third, and fourth resolution layers in the encoder. We show the segmentation results of different variants. Using 3D convolution in the encoder leads to a performance drop compared with depth-wise convolution. Using transformer E2 leads to $73.21\%$, which is better than using transformer E3 or E4. This may due to the reason that identification of cancer areas in pathology images does not require highly semantic information. Thus, the highly semantic information in spatial dimension does not help too much in learning long-range context of the spectrum. Besides, Sparsity and spectral normalization are two vital schemes needed to be learned with transformers to achieve better segmentation performances.

\section{Conclusion}
In this paper, we present Spectral Transformer (SpecTr) for hyperspectral pathology image segmentation, which employs transformers to learn the contextual feature across spectral bands. Two vital schemes are introduced to assist context learning: (1) A sparsity scheme is adopted to learn context-dependent sparsity patterns, and improve the model's performance and interpretability. (2) A spectral normalization method is proposed, to conduct separate normalization for the feature map on each spectral location and eliminate the interference caused by distribution mismatch among spectral images. We evaluated SpecTr on cholangiocarcinoma segmentation dataset. Experiments show the superiority of the proposed method for hyperspectral pathology image segmentation.

\bibliographystyle{splncs04}

\bibliography{mybib}
\end{document}